# An Adaptive Parallel Integrator of Ordinary Differential Equations System for Space Experiment Simulation

Atanas Marinov Atanassov

Solar Terrestrial Influences Institute, Bulgarian Academy of Sciences,
Stara Zagora Department, P.O. Box 73, 6000 Stara Zagora, Bulgaria
E-mail: At_M_Atanassov@yahoo.com

*Abstract*

*Different possible sources are discussed for enhancement of the calculation time when solving ordinary differential equations systems to forecast the space objects motion. This paper presents an approach for building an integrator of ordinary differential equations systems for simultaneous solving of motion equations of multiple objects. A parallelization of calculation on the base of threads is offered. A method for synchronization is presented. The technological advance and the invasion of multi-core processors make actual the examined approach for developing an integrator of ordinary differential equations systems.*

### 1. Introduction

Computer simulation has long ago become a tool for quality improvement of space experiments, which is applied in their design, planning and control. The various dynamical models are based on ordinary differential equations systems (ODEsS). The intensive use of such models (for example, of the motion of different bodies in the around Earth's or other planets' space) may be related with increase of the consumed processor time in the case of computer realization accomplishment. When tackling this issue (if it exist), the focus is on creating more effective numerical methods and computer programs for ODEsS integration.

The delay in the course of processor frequency increase is expected to be successfully compensated with increase of their core number. Presently, the AMD and Intel available double core processors on the market have reasonable prices. Quad core processors are manufactured, too. However, soon expected the manufacturing of single-core processors to end and the development of 8 and 16 core processors is expected. Multi-core processors are intended at disposing two or more processors on one crystal in order to work at lower voltage and with lower frequency. The latter aims to reduce released heat, which is a major issue in increasing operation frequency. This, however, does not ensure automatically raising of the calculation power of multi-core processors, equivalent the one-core processors. It is also necessary to transform serial algorithms into parallel ones, taking into account the specifics of the processor's architectures. The perspective for the personal computers to compete with the expensive computer clusters, specialized in parallel calculations, will require a change in the algorithmic thinking and the development of new program tools on the basis of calculation processes' parallelization.

The integration of ODEsS as an initial value problem (IVP) can be written down in the following way:

$$\dot{y}_i = \varphi_i(t, y_i), \quad y_i(t_0) = y_i^0 \qquad (1)$$

The solution of (1) as IVP is obtained as a series of values of $y(t_m)$ for $t_m = t_0 + m.\Delta t$.

The keystones in the parallel calculations theory within the context of IVP are formulated in a series of articles [1, 2, 3]. Similar issues are examined in [4] with application in the field of orbital dynamics. In [5, 6] the prognostication of the orbital motion of a large number (N $\sim 10^3$) of objects is examined on the basis of analytical methods. [7, 8] expose other approaches to solving the same problem, however, with different numerical methods.

The development of ODEsS integrators (ODEsSI) and their application for integration of space bodies' motion equations is one of the fields, which has always been attractive for its final objective- achieving better precision and effectiveness.

In this work, we will focus on an approach for building a parallel ODEsSI (PODEsSI), based on explicit Runge-Kutta-Fehlberg (RKF) schemes [9, 10] and on its application for satellite orbits' integration. Preliminary results are obtained, using double-core processors. The calculation model's parallelization is based on threads.

**2. Space objects motion equations.**

We can examine the solution of a satellite's motion equations (material objects, N>>1) in the gravitational field of a central body as IVP:

$$\ddot{\vec{r}}_n = \vec{f}_n(t, \vec{r}_n, \dot{\vec{r}}_n), \vec{r}_n(t_0) = \vec{r}_n^0, \dot{\vec{r}}_n(t_0) = \vec{v}_n^0. \qquad (2)$$

Index $n$ is used to denote the object: $1 \leq n \leq N$. System (2) is of the second order and can be reduced to a first order system:

$$\left\| \begin{array}{l} \dot{\vec{r}}_n = v_n \\ \dot{\vec{v}}_n = \vec{f}_n(t, \vec{r}_n, \dot{\vec{r}}_n), \quad \vec{r}_n(t_0) = \vec{r}_0, \quad \dot{\vec{r}}_n(t_0) = \vec{v}_0 \end{array} \right. \qquad (3)$$

The right-hand side of (2) can include in addition to the acceleration from a central gravitational force $\vec{f}_n^{grav}$ different kinds of perturbations [11, 12]:

$$\vec{f}_n(t, \vec{r}_n, \dot{\vec{r}}_n) = \vec{f}_n^{grav} + \vec{f}_n^{atm} + \vec{f}_n^{Sun} + \vec{f}_n^{Moon} + \vec{f}_n^{light} + \vec{f}_n^{el\,din} \qquad (4)$$

By analogy with [1] we can point to several reasons as a result of which the solution of one IVP can be related with the use of much calculation time:
- complex calculation models are used for the calculation of different disturbances in (4) ;
- the mathematical model describes the motion of numerous objects;

- the integration interval [$t_0$,$t_{end}$] is large;
- multiple solutions of IVP to determination of the simulation model's parameters.

The computer modelling of multi-satellite space experiments, involving a large set of instruments, intended to solve multiple scientific problems, aimed at producing numerous parameters, contains the above-mentioned reasons. Appropriate orbital elements are determined during the experiment design stage, which, within the expected instrument operation period, are expected to have optimum conditions for their implementation. The simulation of different aspects of the instruments' operation in model conditions allows specifying some of their parameters (optical, mechanical, electrical, informational, etc.) in order to obtain statistically reliable results.

### 3. Possibilities for parallelization.

Gear [1] points to two approaches for calculation parallelization by numerical integration of ODEsS:
- Parallelism across the method;
- Parallelism across the system.

The phrase „parallelism across the method" expresses the possibility for different calculation stages within the framework of one method to be executed independently and simultaneously on different processors. This paper will be focused mainly on the explicit classical RKF schemes, which are used to build the integrator of ODEsS (OSEsSI). The used schemes feature different precision order and are based on the calculation of functions $\vec{f}_n(t_k, \vec{r}_n, \dot{\vec{r}}_n)$ for $t_m < t_k < t_{m+1}$. Each of the stages, related with calculation of $g_{i,0}$ and $g_{i,k}$ for moments $t_k$ is based on the previous one: $\vec{f}_n(t_{k-1}, \vec{r}_{n,k-1}, \dot{\vec{r}}_{n,k-1})$ and $g_{i,k-1}$. As a result, the possibilities for parallel calculations of coefficients $g_{i,k}$ are restricted with regards to the separate system equations. Each of the six $g_{i,k}$ can be calculated on a separate processor; with two processors, each of them can be computed by three coefficients $g_{i,k}$. The implementation of such kind of parallel calculations, however, requires special compilers.

The phrase "parallelism across the system" means that one equations or a group of equations, part of a large ODEsS, can be solved on a separate processor. This kind of parallelism reflects to some extent the character of the solved problem. It is very suitable for application in simultaneous integration of a large number of equations of type (2).

The parallelization of adaptive integrator of ODEsS, based on "parallelism across the system" is examined in this paper.

### 4. Multi body ODEsS integrator.

Fig. 1 shows the functional diagram of the integrator. The basic subroutine ***rkfasd*** controls the choice of integration scheme. The classical schemes of Fehlberg [9,10] are used – subroutines ***prkf0a, prkf2a, prkf4a, prkf6a*** and ***prkf8a***. Subroutine ***kalkgr*** serves to evaluate the error and to verify whether the current scheme is suitable or, another should be selected.

Integration with a variable step within interval Δt is selected, if the scheme with maximal precision is not sufficient. Subroutine ***pertur*** calculates the right-hand side of (3). The serial version of the integrator is directly called on the basis of subroutine ***rkfasd***. It is designed as storage automata and can integrate simultaneously many ODEsS with individual scheme choice for each system or step-size control, if the precision of the highest-order scheme is not sufficient.

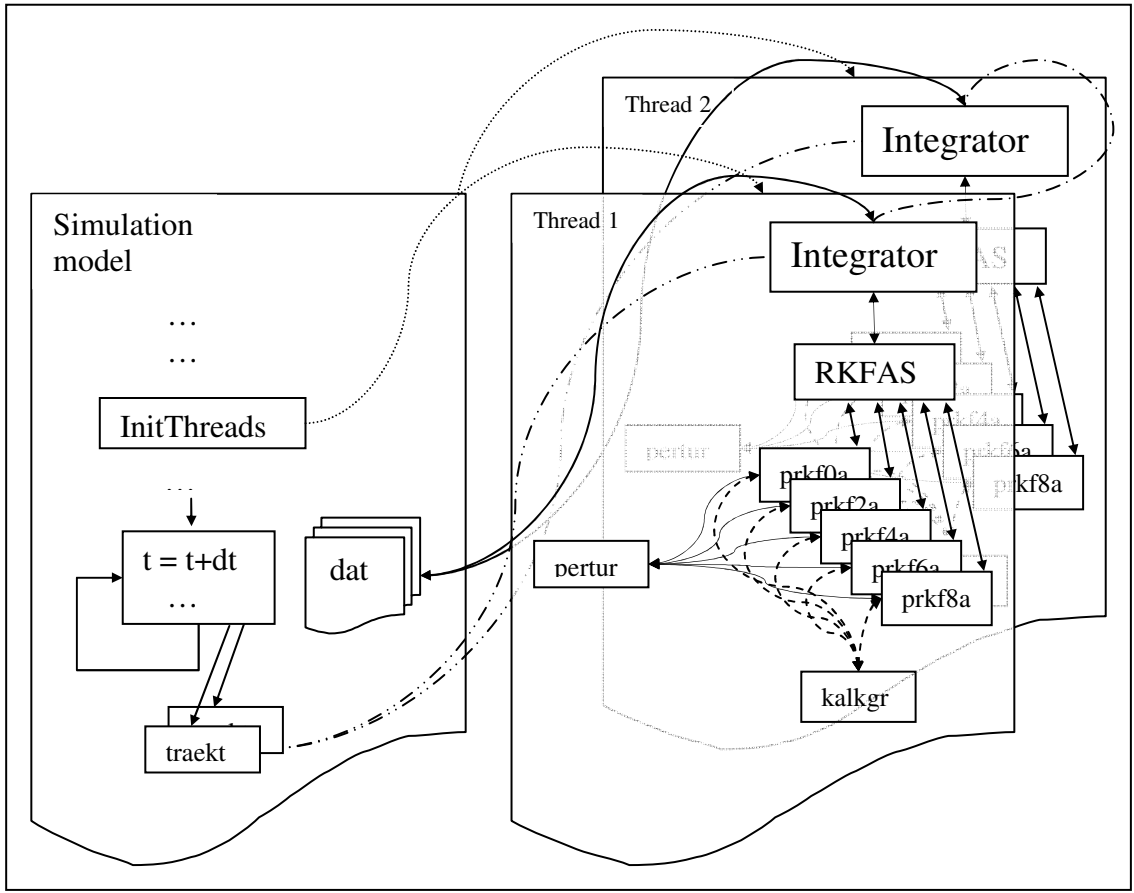

Fig. 1. General functional diagram of the integrator.

The parallel version of ODEsSI (PODEsSI) is initialized before the first integration step. A certain amount of integrator threads (Ith) are started according to the number of processor cores. These threads are left in suspended state up to the first addressing o PODEsSI. The ITh perform the selection of different ODEsS in competitive conditions on the basis of a synchronizing code, shown below and proceeds until reaching the maximum number. The choice is protected from doubling of ODEsS and solving one system more than once from different IThs on the basis of a special event. While one ITh is performing the choice of ODEsS, the remaining threads are waiting for the choice to be completed. When one ITh finishes the integration of the respective system for the current moment $t$ and step $\Delta t$, it tries to select the next system, if there are any remaining.

## 5. Thread's synchronization.

All calculations, related with the integration of ODEsS are portioned and each portion is related with one specific time step. This requires synchronization of the main and secondary threads ITh. The respective code in the basic parent's thread is, as follows:

```
SUBROUTINE  traekt(num_sat,dt)
           ...
  glb_counter = 0
a:DO  i=1,num_th
     k= SetEvent(thread_par(i)%ha_beg) !  Events for releasing threads
  END DO a
     k= WaitForMultipleObjects(num_th, ha_end,WaitAll,Wait_infinite)
b:DO  i=1,num_th
     k= ResetEvent(thread_par(i)%ha_end) ! Events for starting threads
  END DO b
           ...
END SUBROUTINE  traekt
```

The above fragment of subroutine **traekt** illustrates the control of the PODEsSI. The counter ***glb_counter*** which counts the number of ODEsS is zeroed before starting the integration. In the first **a:DO** construct, the event objects ***ha_beg*** for starting the threads' operation are set out in a signaled state. After this subroutine **traekt** passes to a waiting state until all treads ascertain that all systems are solved. The second **b:DO** construct makes non-signaled the event objects ***ha_end*** which point to the end of each ITh. The latter ensures the waiting state in the integration step.

Instead of the traditional call of the integrator as a subroutine, by analogy with the serial version without parallel calculations (connection by address) we have now synchronization of the parent's thread and the IThs. Instead of transmitting data by actual and formal arguments, now global data are used for transferring the address of allocatable user-defined type data which contain the generic data- coordinates of radius vectors, velocities, tolerance, etc.

The threads IThs perform two kinds of synchronizations. The first one is with the basic thread and should perform the step integration of ODEsS. For this purpose, event objects with handlers ***ha_beg*** are used to start the choice of ODEsS and their solving; other event objects with handlers ***ha_end*** (by one for every ITh), signal the end of the integration. These event objects are the same as the above described in connection with subroutine **traekt**. The second communication is between IThs only and is devoted to concurrent choice for ODEsS. As a result of this synchronization it is possible that while one thread is choosing a subsequent ODEsS, all others are waiting. In this way, the solution of one ODEsS with different threads is avoided. The allocation of the systems among IThs is achieved within the frames of "**b: DO WHILE()**" construction, shown in the next fragment:

```
  SUBROUTINE    Integrator (th_id_num)
    ...
a: DO WHILE(.true.)
       k= WaitForSingleObject(ha_beg,WAIT_INFINITE) ! Event for threads starting
   b: DO WHILE(glb_counter.LT. numsat)              ! ODEsS concurrent distribution
        k= WaitForSingleObject(ha_1,WAIT_INFINITE)  ! prohibition of ODEsS distribution
              glb_counter= glb_counter + 1;          ! serial number of ODEsS
              loc_counter= glb_counter;              ! remembering in local thread's storage
          k= SetEvent(ha_1)                          ! release of ODEsS distribution
            IF(loc_counter.GT.numsat) EXIT
         CALL  rkfasd(loc_counter,m,transfer_data(loc_counter)%t , ... )
       END DO b                                     ! end of concurrent distribution
    k= ResetEvent(thread_par(th_id_num)%ha_beg)
    k=    SetEvent(thread_par(th_id_num)%ha_end)
  END DO a
   ...
END SUBROUTINE  Integrator
```

## 6. Analysis of effectiveness.

The entire PODEsSI code is developed on Compaq Visual Fortran Professional Edition v.6.6 under WindowsXP, using QuickWin library. Tests were conducted on one core Intel and AMD processors for checking the correctness. Experiments were conducted on a double core Athlon to establish the effectiveness of the integration of motion equations of six satellites with different orbits. The experiments have shown that, with double core processors, the calculation time is not possible to be decreased. The serial code execution time which is transformed into a thread increases about twice. Besides, time is necessary for the threads' synchronization between parent's and children's IThs. As a result, the execution time of the parallel version exceeds the time of the serial one by about 30%.

## 7. Conclusion and future work.

The basic part of the theoretic and experimental investigations which are related with different aspects of solving ODEsS-IVP has private character and restricted practical application. The increase of the effectiveness and the investigation of ODEsS integration have major significance in solving complex problems related with difficult simulation models of physical reality. These models can describe both moving satellites and different satellite subsystems. The whole simulation model can include models of instruments and mechanisms, of the environment where the measurements are conducted, as well as of investigation method and the investigated object. Some of the private models can be deterministic and others - stochastic. In such conditions both the IThs and other threads can be initialized in completely competitive conditions- for example, for calculation of the geophysical parameters by orbits of satellites included in the model, the geomagnetic field, the directions of the instrument optical axis, for measurements imitation and analysis of the statistical reliability. Further analysis is necessary to explain the behavior and effectiveness of the adaptive parallel integrator in the framework of such a global model.

We can consider normal the above results obtained using double core processors. It is necessary to conduct experiments with quad core processors. We can expect decrease of the execution time of a parallel code compared to the execution time of a series code.

**Reference:**
1. Gear C. W., The Potential for Parallelism in Ordinary Differential Equations, in Computational Mathematics, ed. Simeon Fatunla, Boole Press, Dublin, pp33-48, 1987
2. K. R. Jackson., A Survey of Parallel Numerical Methods for Initial Value Problems for Ordinary Differential Equations, IEEE Transactions on Magnetics, 27 (1991), pp. 3792-3797.
3. L. C. Stone, S. B. Shukla, B. Neta, Parallel Satellite Orbit Prediction Using a Workstation Cluster, *International J. Computer and Mathematics with Applications*, **28**, (1994), 1–8.
4. B. Neta, Parallel Solution of Initial Value Problems, *Proc. Fourth International Colloquium on Differential Equations*, D. Bainov, V. Covachev, A. Dishliev (eds), Plovdiv, Bulgaria, 18-23 August 1993, **2**, (1993), 19–42.
5. W. E. Phipps, B. Neta, D. A. Danielson, Parallelization of the Naval Space Surveillance Satellite Motion Model, *J. Astronautical Sciences*, **41**, (1993), 207–216.
6. Beny Neta, D. A. Danielson, Sara Ostrom, Susan K. Brewer, Performance of Analytic Orbit Propagators on a Hypercube and a Workstation Cluster.
7. Atanassov At., Integration of the Equation of the Artificial Earth's Satellites Motion with Selection of Runge-Kutta-Fehlberg schemes of Optimum Precision Order., Aerospace Research in Bulgaria, 2007, 21, 24-34.
8. Kolyuka Yu.F., T.I. Afanasieva, T.A. Gridchin, Precise Long-Term Prediction of the Space Debris Object Motion, Proceedings of the 4th European Conference on Space Debris, 18-20 April 2005, ESA/ESOC, Darmstadt, Germany.
9. Fehlberg E., Klassische Runge-Kutta formeln funfter und siebenter Ordnung mit Schrittweitenkontrolle., Computing, v. 4, p.93-106, 2969.
10. Fehlberg E., Klassische Runge-Kutta formeln vierter und nidrigerer Ordnung mit Schrittweitenkontrolle und ihre Anwendung auf Warmeleitungsprobleme., Computing, v. 6, p. 61-71, 1970.
11. В. И. Прохоренко. Описание универсальной программы расчета навигационной информации о положении искусственного спутника Земли. Пр. ИКИ АН СССР № 263, 1976, с.80.
12. П. Р. Попович, Б. Ц. Скребушевский, Баллистическое проектирование космических систем., Москва, Машиностроение, 1987, 239с.